\documentclass[11pt, letterpaper, oneside, article]{memoir}
\settrimmedsize{11in}{8.5in}{*}

\settrims{0in}{0in}

\settypeblocksize{9in}{6.5in}{*}

\setlrmargins{1in}{*}{*}

\setulmargins{1in}{*}{*}
\checkandfixthelayout
\usepackage[version=3]{mhchem}
\usepackage[sc]{mathpazo}
\usepackage[T1]{fontenc}
\usepackage{subfig}
\usepackage{graphicx}
\usepackage{amsmath}

\usepackage[protrusion=true,expansion=true]{microtype}
\captionnamefont{\normalfont\small}
\captiontitlefont{\normalfont\small}
\usepackage{authblk}
  \usepackage[hyperfootnotes = true]{hyperref}
  \hypersetup{
    colorlinks = true,
  	citecolor=blue,
    urlcolor = blue,
    linkcolor=blue,
    pdfpagemode = UseNone
  }
  
\begin{document}
\title{Realization of ``Trapped Rainbow'' in 1D slab waveguide with Surface Dispersion Engineering}
\author{Rui Yang}
\author{Wenkan Zhu}
\author{Jingjing Li}
\affil{Department of Electrical and Computer Engineering, University of Illinois at Chicago, USA
\\ \href{mailto:jili@uic.edu}{jili@uic.edu}}
\maketitle
\begin{abstract}
We present a design of a one dimensional dielectric waveguide that can trap a broad band
light pulse with different frequency component stored at different positions, effectively forming a
``trapped rainbow''\cite{tsakmakidis_trapped_2007}. The spectrum of
the rainbow covers the whole visible range. 
To do this, we first show that the dispersion of a \ce{SiO2} waveguide 
with a Si grating placed on top can be engineered by the design
parameter of the grating.  Specifically, guided modes with zero group 
velocity(frozen modes) can be realized. 
Negative Goos-H\"anchen shift along the surface of the grating is
responsible for such a dispersion control. 
The frequency of the frozen mode is tuned by changing the lateral
feature parameters (period and duty cycle) of the grating.
By tuning the grating feature point by point along the waveguide, 
a light pulse can be trapped with different frequency
components frozen at different positions, so that a ``rainbow'' 
is formed.
The device is expected to have extremely low ohmic loss because
only dielectric materials are used.  A planar geometry also promises
much reduced fabrication difficulty.
\end{abstract}

\section{Introduction}
To use photons as the information carrier is currently under intense study because of 
the almost unlimited bandwidth and the high energy efficiency.  Many 
devices in an optical network depend on the capability to control the 
dispersion property of a waveguide.  One example is the slow light for which
the group velocity of light is much smaller than that in the free space.
Applications of slow light devices include optical buffers, nonlinear optics, 
and optical signal processing.  Whereas the slow light is usually realized on 
Bose-Einstein condenstate\cite{hau_light_1999},
to build solid-state slow light devices is of great practical importance.  The former 
is based on electromagnetically induced transparency (EIT) and usually demands on
bulky, ultra-low temperature apparatus, while the latter is of much lighter weight
and is applicable for on-chip integration.  
Photonic crystal is used almost exclusively for on-chip slow light devices.  
Usually, the part of the dispersion curve around the edge of the 
reduced Brillouin zone is used.  This part of the dispersion curve are flattened 
because of the coupling between the forward and backward propagating modes, and exhibits 
a very small group velocity\cite{ma_wave_2013, li_systematic_2008, schulz_dispersion_2010,
colman_control_2012}. In 2007, a new idea to realize slow and stopped light 
was proposed in theory by Tsakmakidis,
Boardman and Hess, making use of the anomalous property of metamaterials\cite{tsakmakidis_trapped_2007}.  In their proposal, the negative 
Goos-H\"anchen shift on a interface between a metamaterial
and a regular dielectric is used.  It was shown that, for a waveguide made
of a metamaterial, when the Goos-H\"anchen shift on the side walls of the waveguide compensates
completely the forward-leap of the ray in a round trip, the guided mode would become,
intuitively, ``frozen'' on the waveguide and no forward power propagation can
be observed.  This is actually a description of slow light using the ray picture. 
Further, a scheme of trapping optical signal of a broad frequency band is proposed: since 
the operating frequency to freeze the light is related to the waveguide thickness, a waveguide segment
of tapered thickness should be able to trap light of a continuous spectrum at different
positions along the waveguide, forming a ``trapped rainbow''.

The idea has since attracted many researchers and different designs have been tried.  
However, the experimental realization of the original idea of ``trapped rainbow'' 
faces great challenge.  In the original design, the metamaterial was treated as a 
homogeneous medium similar to a regular 
dielectric, while in reality such an artificial material is always composed of 
discrete inclusions with strong temporal and spatial dispersion.  Up to now,
the best optical metamaterial uses inclusions of $\sim \lambda_0/3$ in size, where
$\lambda_0$ is the free space wavelength at the operating frequency.  When used
to build a waveguide which itself might only be a few wavelengths in width, 
thus modeling the waveguide as a homogeneous one is problematic.  The optical
metamaterial usually operates at a frequency up to the near infrared.  Little
progress has been made for metamaterials working in the visible band with
reasonably good property.  Also, metamaterials are inevitably dispersive,
and there have been no report on metamaterials with negative $\epsilon$
and 
 $\mu$ that cover the whole visible domain.
Further, the ohmic loss related with metamaterial is a formidable factor.  In the optical
frequency domain, plasmonic materials (gold or silver) are used almost exclusively
to build metamaterials.  Their ohmic loss is far from tolerable for 
the application of ``trapped rainbow'', and might erase any feature related to
the broad band rainbow trapping.  There have been a few reports on the experimental 
demonstration of the ``trapped rainbow'' after the theoretical
proposal\cite{gan_experimental_2011, gan_surface_2011,smolyaninova_experimental_2010}.  However,
none has actually used the approach proposed in the original paper that was based on the
negative Goos-H\"anchen shift.  Rather,
they are realized on the edge of the Brillouin zone of a plasmonic periodic structure.
Inevitably, the strong ohmic loss makes the trapping effect very weak.

In this paper, we numerically demonstrate an approach to realize frozen mode based
on the negative Goos-H\"anchen shift.  The proposed approach uses only dielectric materials, 
thus could have extremely low ohmic loss.  The ``trapped rainbow'' is then
realized by a waveguide with chirped or adiabatically tuned design of the frozen mode
waveguide.  Our approach is the first, complete demonstration for frozen mode and trapped rainbow
that uses negative Goos-H\"anchen shift, the original idea in Ref.\cite{tsakmakidis_trapped_2007}.  In the
following, we first review the idea of using negative Goos-H\"achen shift to construct
a frozen mode, and our recent discovery that negative Goos-H\"anchen shift, sometimes of giant
magnitude, can be realized on the surface of a dielectric decorated by a grating.  
We then demonstrate rigorously that a ``frozen mode'' where the Goos-H\"anchen shift 
fully compensates the forward leap of the ray in a round trip inside a slab waveguide 
indeed corresponds to zero group velocity of the guided mode, and show 
numerical results of such a frozen mode.  Based on this waveguide that supports frozen mode,
we then show our designs of ``trapped rainbow'', where a broadband
pulse is stopped with different frequency components trapped on different positions
along the device.

\section{Goos-H\"anchen shift and frozen mode}
When a Gaussian beam is totally reflected from a surface, the axis of the reflected
beam experiences a lateral shift with respect to the position predicted by geometric
optics\cite{Goos_1947}.  Such a phenomenon is named after the discoverers Goos and 
H\"anchen, and has been shown as an
example of discrepancy between geometric optics and the wave nature
of the light.  The shift is related to the change of the reflection phase for
different plane wave components of the incident beam.  Mathematically, 
the Goos-H\"anchen shift can be evaluated as\cite{Tamir_Integrated_1979}
\begin{equation}
\label{eqn:GHS}
z_s = -\frac{\partial\phi}{\partial k_x}
\end{equation} 
where $\phi$ is the phase of the reflection coefficient of the plane 
wave component with a lateral wavenumber of $k_x$. Here an $e^{-i\omega t}$
time variation is assumed for the electromagnetic field.  
The Goos-H\"anchen shift is usually positive for the reflection from an interface
between two regular dielectrics, while negative on interfaces 
between regular dielectrics and plasmonic material or metamaterial.

The negative Goos-H\"anchen shift has attracted a lot of research interest,
one of which is to control the direction of energy flow in a dielectric slab waveguide 
with respect to the wave vector of the guided mode, as discussed in
the same paper that proposed the trapped rainbow\cite{tsakmakidis_trapped_2007}.
Whereas the original discussion was from a rather intuitive approach, here we would like 
to give a rigorous mathematical description.  Considering a slab waveguide made of a 
dielectric of refractive index $n$ and thickness $h$.  A guided mode
can be described as a plane wave total-internally reflected back and forth on the two 
interfaces that satisfies the following relationship:
\begin{equation}
\label{eqn:disp}
2 n h k_0 \cos\theta + \phi_1(\theta, k_0) + \phi_2(\theta, k_0) = 2m\pi
\end{equation}
where $k_0$ is the free space wavenumber, $\theta$ is the angle of incidence,
$\phi_j(\theta, k_0)$, $j = 1, 2$ is the phase loss (the phase
of the reflection coefficient) on the two side walls, respectively, while $m$
an integer.  When written in terms of $k_x$, the wavenumber parallel to the
waveguide wall, we have
\begin{equation}
2 h \sqrt{n^2 k_0^2 - k_x^2} + \phi_1(k_x, k_0) + \phi_2(k_x, k_0) = 2m\pi
\end{equation}
Take the total differential of both sides with respect to $k_0$ and $k_x$, we get{\small
\begin{equation}
\label{eqn:total_diff}
2 n h \frac{\Delta k_0}{\sqrt{1 - k_x^2 / (n k_0)^2}} 
  - 2 h \frac{k_x}{\sqrt{n^2 k_0^2 - k_x^2}}\Delta k_x
  +\sum\limits_{j = 1 ,2}\left(\frac{\partial \phi_j}{\partial k_0}\Delta k_0
  + \frac{\partial \phi_j}{\partial k_x}\Delta k_x \right) = 0
\end{equation}
}
which is a relationship a guided mode must satisfy in addition to \eqref{eqn:disp}.
When deriving the former equation, we assume that $n$ does not change with frequency, which
is a reasonable assumption for dielectric waveguides.
Divide both sides by $\Delta k_x$ and take the limit of 
$\Delta k_x\rightarrow 0$,  we get
{\small
\begin{equation}
\label{eqn:zgv_GHS}
\left(\frac{2 n h }{\sqrt{1 - k_x^2 / (n k_0)^2}} 
  + \sum\limits_{j = 1 ,2}\frac{\partial \phi_j}{\partial 
  k_0}\right)\frac{\partial k_0}{\partial k_x} 
  = 2 h \frac{k_x}{\sqrt{n^2 k_0^2 - k_x^2}} 
  + \sum\limits_{j = 1 ,2}-\frac{\partial \phi_j}{\partial k_x}
\end{equation}
}

\begin{figure}[htbp]
\hspace*{\fill}
\subfloat[]{ %
\label{fig:posGHS_posSz}
\includegraphics[width=0.47\textwidth]{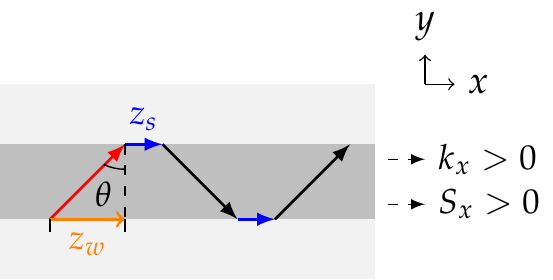}
}
\hfill
\subfloat[]{ %
\label{fig:negGHS_posSz}
\includegraphics[width=0.47\textwidth]{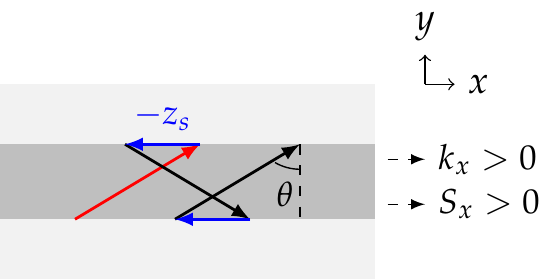}
}
\hspace*{\fill}
\\
\hspace*{\fill}
\subfloat[]{ %
\label{fig:negGHS_noSz}
\includegraphics[width=0.47\textwidth]{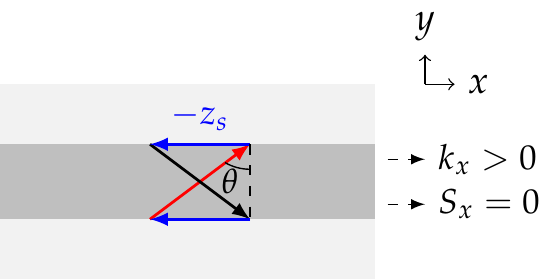}
}
\hfill
\centering
\subfloat[]{ %
\label{fig:negGHS_negSz}
\includegraphics[width=0.47\textwidth]{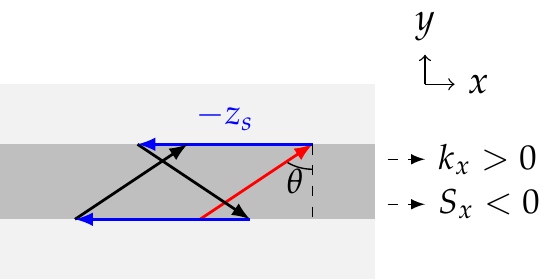}
}
\hspace*{\fill}
\caption{\label{fig:GHS_waveguide} Energy flow and Goos-H\"anchen shift in a planar waveguide
. The incidence is colored in red, Goos-H\"anchen shift $z_s$ in blue, while the forward displacement 
$z_w$ in orange. The rays travels inside the waveguide are in black.}
\end{figure}
Notice that $\partial\phi_i / \partial k_0$ is always positive. 
This is because $\partial\phi_i / \partial k_0$ is the delay of the center
of the Gaussian pulse at the reflection of the interface.  For lossless
reflection (which is the case here),
this delay must be positive for a causal system.  This means the sign
of the right hand side completely determines the sign of 
$\partial k_0 / \partial k_x$, which is proportional to the
group velocity.  For the right hand side, 
the second term is the Goos-H\"anchen shift on the two side walls.  
Also notice that $k_x/\sqrt{n^2 k_0^2 - k_x^2} = k_x / k_y$. Thus, if
we let $z_w = h k_x/\sqrt{n^2 k_0^2 - k_x^2}$, $z_w$
is actually the forward displacement of the ray when propagating from
one side wall to the other (see Fig.\ref{fig:posGHS_posSz}).  
Combining the contribution of $z_s$ and $z_w$ together, the right hand side
of \eqref{eqn:zgv_GHS} gives the total $x$ direction displacement of 
a ray in a round trip, as we see in Fig.~\ref{fig:GHS_waveguide}.  
When the waveguide and the surrounding medium are both made of regular 
dielectrics, the Goos-H\"anchen shift is positive, thus the right hand side
is always positive.  This means we always have $\partial k_0 / \partial k_x > 0$.
Things become interesting when we have negative Goos-H\"anchen shift
on one or both of the side walls, especially when it is of large
magnitude so that the total displacement is
negative (Fig.~\ref{fig:negGHS_negSz}).  In this case, the group velocity would be
negative, and the energy propagates anti-parallel to $k_x$.  
When the Goos-H\"anchen shift is just enough to make the right hand side 
goes to zero (the situation described by Fig.~\ref{fig:negGHS_noSz}), 
we have $\partial k_0 / \partial k_x = 0$, and a 
``frozen mode'' of the waveguide is formed.  This is a guided mode
with finite propagating constant, but zero net power propagation. All
these conclusions are consistent with those in Ref.\cite{tsakmakidis_trapped_2007}
but with rigorous mathematical analysis.  We would like to point
out that the conclusions only hold when the waveguide material has no
temporal or spatial dispersion, i.e. $\partial n/\partial k_0 = 0$ and
$\partial n / \partial k_x = 0$, as assumed when deriving \eqref{eqn:total_diff}.
This is a reasonable assumption for dielectric waveguide, but not for
metamaterial waveguides.  

\section{Negative Goos-H\"anchen shift and frozen mode on a dielectric grating}
\begin{figure}[hbt]
\centering
\hspace*{\fill}
\begin{minipage}[c]{0.48\textwidth}
\centering
\subfloat[]{ %
\label{fig:grating_schem}
\includegraphics[width=\textwidth]{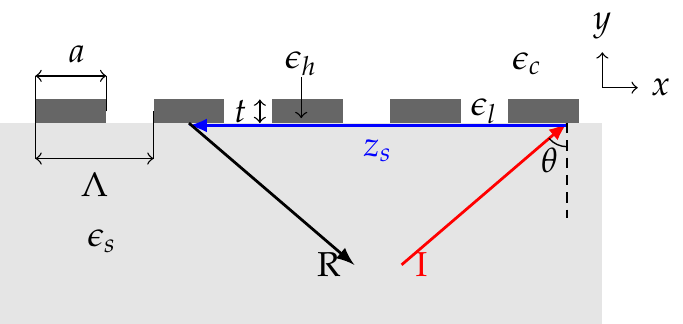}
}
\end{minipage}
\hfill
\begin{minipage}[c]{0.48\textwidth}
\centering
\subfloat[]{ %
\label{fig:phase_vs_theta}
\includegraphics[width=\textwidth]{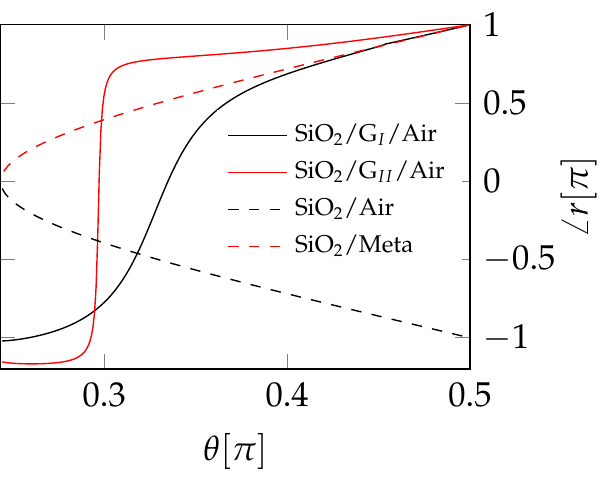}
}
\end{minipage}
\hspace*{\fill}
\caption{\protect\subref{fig:grating_schem}: Schematic of an infinite thick substrate decorated 
by a dielectric grating. I and R stand for incidence and reflected beam respectively. $z_s$ is the Goos-H\"anchen shift.
Duty cycle is defined as $\Gamma = a /\Lambda$.
\protect\subref{fig:phase_vs_theta}:Reflection phase vs Incidence angle $\theta$ of different interfaces
when incidence coming 
from the \ce{SiO2} substrate. Operating free space wavelength is $1.5\mu m$. Grating$_{I}$ 
parameters: $\Lambda_I=0.53\mu m$, $t_I=0.097\mu m$,
$\Gamma_I=0.65$. Grating$_{II}$ parameters:$\Lambda_{II}=0.43\mu m$, $t_{II}=0.11\mu m$,$\Gamma_{II}=0.93$.}
\end{figure}

A negative Goos-H\"anchen shift is crucial in building a frozen mode. 
This can be achieved on the surface of plasmonic materials or metamaterials,
but is usually accompanied with large ohmic losses.  However, it is possible to make negative
Goos-H\"anchen shift using completely dielectric devices, as we demonstrated in
a recent publication\cite{yang_giant_2014}.  The system under consideration is 
shown in Fig.~\ref{fig:grating_schem}, where a thin grating made of Si
is placed on a substrate of \ce{SiO2}.  For certain grating design,
the phase of the reflection coefficient for incidence from the
\ce{SiO2} side is of very different nature compared to that on the 
\ce{SiO2}/Air interface, as we see in Fig.~\ref{fig:phase_vs_theta}
in which $S$ polarized incidence is studied, i.e. $E_z$ is the only electric field component.
Whereas the phase decreases with the incident angle for a \ce{SiO2}/Air interface
indicating a positive Goos-H\"anchen shift 
(see \texttt{-{}-{}-} 
in Fig.~\ref{fig:phase_vs_theta}),
on the \ce{SiO2}/Grating interface, the phase increases, exhibiting
a negative Goos-H\"anchen shift 
(see {\color{red}---} and {\color{black}---} 
in Fig.~\ref{fig:phase_vs_theta}).
This is similar to that of a \ce{SiO2}/Metamaterial case
(see {\color{red}\texttt{-{}-{}-}} 
in Fig.~\ref{fig:phase_vs_theta}).
The negative Goos-H\"anchen shift is related to the guided mode of the
grating.  For the second band of the guided mode 
of the grating, the energy propagates to the opposite direction
of the wave vector.  The part of the dispersion curve for this band 
that is between the light lines of the free space and the substrate is 
leaky on the substrate side, and can couple to the incident beam efficiently.  
According to a commonly accepted explanation, the negative energy propagation
with respect to the lateral wave propagating direction is responsible for
the negative Goos-H\"anchen shift\cite{yang_giant_2014, Renard_Total_1964}.
The amount of Goos-H\"anchen shift can be controlled by the 
grating design: depending on the parameters of the grating, we may have
a very large (a steep $\phi$-$k_x$ curve) or a mediocre (a slow-varying 
$\phi$-$k_x$ curve) Goos-H\"anchen shift.  In fact, the amount of Goos-H\"anchen shift
ranges from tens of nanometers up to several millimeters.  In our study,
a Goos-H\"anchen shift of more than $5000$ times of the free space
wavelength\cite{yang_giant_2014} has been observed. 

\begin{figure}[!hbt]
\hspace*{\fill}
\begin{minipage}[b]{0.58\textwidth}
\centering
\subfloat[]{ %
\label{fig:disp}
\includegraphics[width=\textwidth]{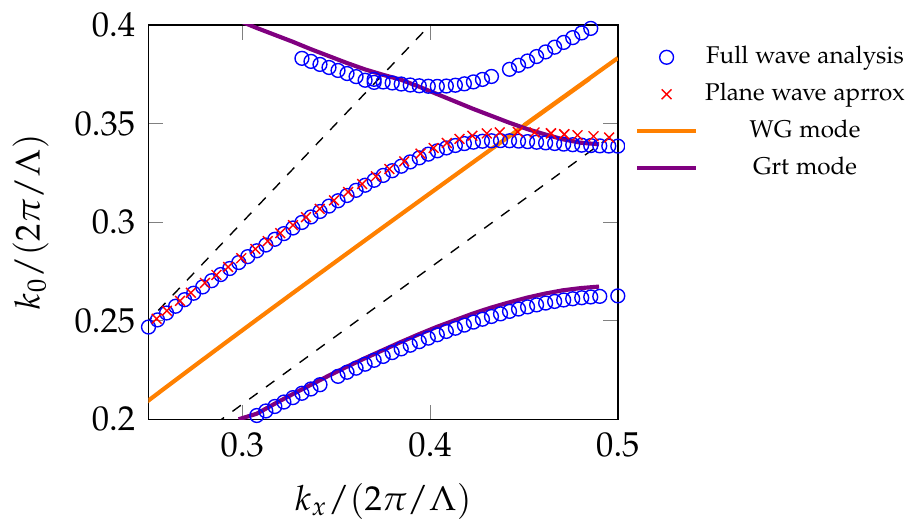}
}
\end{minipage}
\hfill
\begin{minipage}[b]{0.4\textwidth}
\centering
\subfloat[]{ %
\label{fig:fields}
\includegraphics[width=\textwidth]{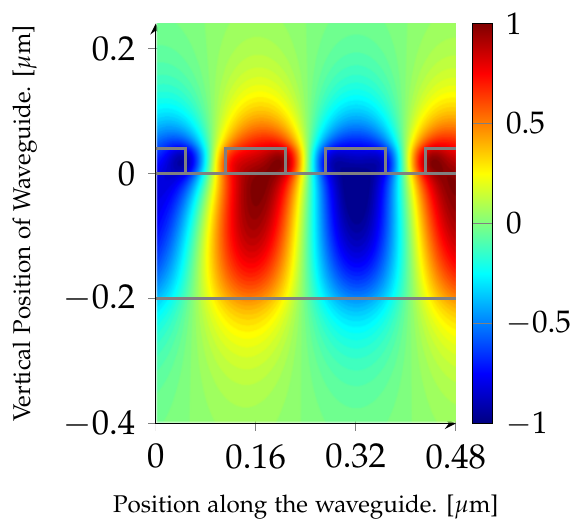}
}
\end{minipage}
\hspace*{\fill}
\caption{\protect\subref{fig:disp}: The dispersion property of waveguide with 
grating parameters: $t=0.04\mu m$,$\Lambda=0.16\mu m$, $\Gamma=0.6$, $\epsilon_h=10.24$,
$\epsilon_l=\epsilon_c=1$; 
waveguide parameters: $h=0.2\mu m$, $\epsilon_s=2.09$. The upper and lower 
dashed lines are light
lines of the free space and the waveguide. 
The blue circles stands for the 
FDTD result while red cross stands for the plane wave approximation analysis.
Solid violet line shows the dispersion of single grating laying on a $\text{SiO}_2$ substrate, while 
soild orange line shows the dispersion of bare waveguide.
\protect\subref{fig:fields}: The instantaneous field distribution of the frozen mode
in three periods of the waveguide. 
}
\end{figure}

With the help of the negative Goos-H\"anchen shift on the grating,
we can readily realize the frozen mode discussed in the former section,
by placing the grating on the sides of a dielectric waveguide.
It turns out that grating on one side is enough to 
realize our goal.  One of the designs makes use of a \ce{SiO2}
waveguide of $200\text{nm}$ thick and a grating of $40\text{nm}$
in thickness.  To calculate the dispersion curve of this grating-decorated
waveguide, we first find out the reflection phase $\phi(k_0, k_x)$
on the \ce{SiO2}/Grating interface and the \ce{SiO2}/Air interface, respectively.
These values are then used in \eqref{eqn:disp} to get the dispersion
relation.  The results are shown in Fig.~\ref{fig:disp} as red cross.
Notice how the dispersion curve bends to form a local extreme where
$\partial k_0 / \partial k_x$ goes to zero at which a frozen mode is formed.
Calculation confirms that the right hand side of \eqref{eqn:zgv_GHS} 
indeed goes to zero at the top of the dispersion curve, which demonstrates
the application of \eqref{eqn:zgv_GHS} in finding a frozen mode.  
The numerical evaluation also shows that the right hand side of \eqref{eqn:zgv_GHS} 
is positive for the part of the dispersion curve with $k_x$ smaller than that
at the top point, and negative when $k_x$ is larger than that at the top. 
This is consistent with the positive or negative group velocity the dispersion
curve shows (see Fig.~\ref{fig:GHS_waveguide}). 
The nature of the negative Goos-H\"anchen shift can be used to understand
the frozen mode.  Recall that 
the negative Goos-H\"anchen shift is usually explained\cite{Renard_Total_1964} 
by an energy flow beyond the reflection interface that is opposite to $k_x$.  
If this power flow compensates completely the forward power
flow inside the waveguide, no net power flow is carried by the 
guided mode, and a zero group velocity is expected.

\begin{figure}[htbp]
\centering
\includegraphics[width=0.6\linewidth]{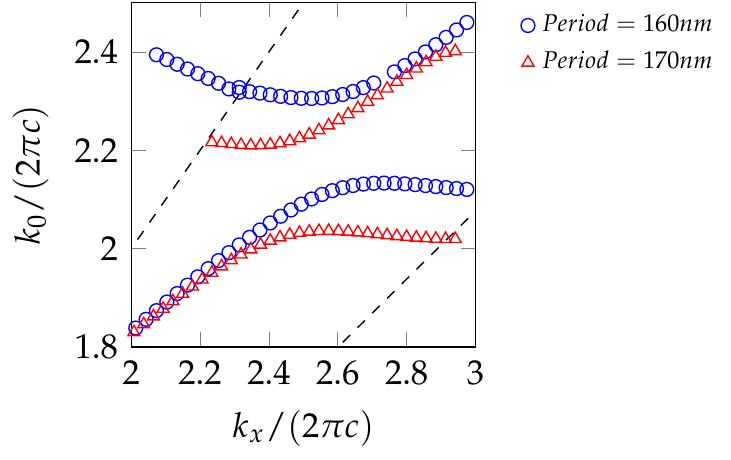}
\caption{The dispersion 
 of two types of waveguide in the difference of $Period=0.16\mu m$(Blue circle) and 
$Period=0.17\mu m$(Red triangle).}
\label{fig:disp_cmpr}
\end{figure}

The evaluation of the dispersion relation using \eqref{eqn:disp}
ignores the high order spatial harmonics of the field around the grating, of course.
To see if this poses any important influence, we also evaluated the 
dispersion relation of the waveguide using full wave analysis. To do
this, we use MEEP, an open source numerical electromagnetic package
based on the finite-difference, time-domain (FDTD) method.  The result
is shown as circles in the same plot of Fig.~\ref{fig:disp}, together
with the guided modes of the same Si grating sitting on a \ce{SiO2}
substrate of infinite thickness.
We can identify the nature of each part of the dispersion curve 
by examining the field distribution of the guided modes.
For the lowest band below the light line of \ce{SiO2}, the electromagnetic field 
is well confined inside the grating, and the dispersion curve overlaps 
well with the guided mode of the grating on \ce{SiO2} substrate.  
These modes are below the light line of \ce{SiO2} and does not couple well with
the propagating plane waves in the \ce{SiO2} waveguide, thus can
not be predicted by \eqref{eqn:disp}.  Rather, this is the guided mode of
the grating itself.  The second band is above the light line of \ce{SiO2}, thus 
the plane waves inside the \ce{SiO2} slab waveguide take part in the formation
of this band.  Notice that the dispersion curve calculated from 
\eqref{eqn:disp} (plotted as 
{\color{red} \tiny $\times$})
indeed overlaps well with that calculated from full-wave analysis.
This means the high order spatial harmonics of the field do not contribute obviously
in forming the mode, and to use \eqref{eqn:disp} for mode calculation is safe. 
The shape of the lower part of the second band is similar
to the dispersion curve of a bare \ce{SiO2} slab waveguide of the same thickness,
but shifted in $k_x$ because of the changed reflection phase on the \ce{SiO2}/grating
wall (see \eqref{eqn:disp}).  
As the frequency increases, the waveguide mode gets close
to the second band of the grating's guided mode where the two 
anti-cross each other, causing the opening of a bandgap.  
Notice that, the second band of the grating mode is also where negative Goos-H\"anchen
shift is observed\cite{yang_giant_2014}.  
The instantaneous field distribution of the frozen mode,
i.e. the mode at the top of the lower band, is shown in 
Fig.~\ref{fig:fields}.  The Poynting vector evaluated from these
simulation results indeed confirm the zero net power flow of this mode.
One important feature needed for a trapped rainbow is the capability 
to tune the frequency of the frozen mode.  In the original paper
of trapped rainbow\cite{tsakmakidis_trapped_2007}, this is realized by
varying the thickness of the waveguide.  
A tapered waveguide requires gray-scale etching,
which is difficult in the conventional micro- and nano- fabrication
developed for planar geometry.  The grating used in our device 
can tune the operating frequency without thickness variation:
we can change the lateral design parameters (the period $\Lambda$ and
the duty cycle $\Gamma$) to modulate the frequency of the frozen mode 
while leaving the waveguide thickness untouched.
It appears that the frozen mode frequency can be varied most effectively
by changing the period.  The effect is shown in Fig.~\ref{fig:disp_cmpr}, 
in which the dispersion curves for two waveguides of the same
\ce{SiO2} slab and grating thickness ($200\text{nm}$ and $40\text{nm}$, 
respectively) but different grating periods (  
{\color{blue} $\circ$}:
$Period = 160\text{nm}$; 
{\color{red} \tiny $\triangle$}: $Period = 170\text{nm}$.)
The result in this plot is again from MEEP simulation.
As we can see, the frequency of the frozen mode (the top of the lower band) is 
obviously changed.  For a period variation of $\sim 6\%$, the frequency is
changed by $\sim 4.5\%$ . 

\section{Trapped Rainbow}
\begin{figure}[hbtp]
\hspace*{\fill}
\begin{minipage}[c]{0.49\textwidth}
\subfloat[]{ %
\includegraphics[width=\textwidth]{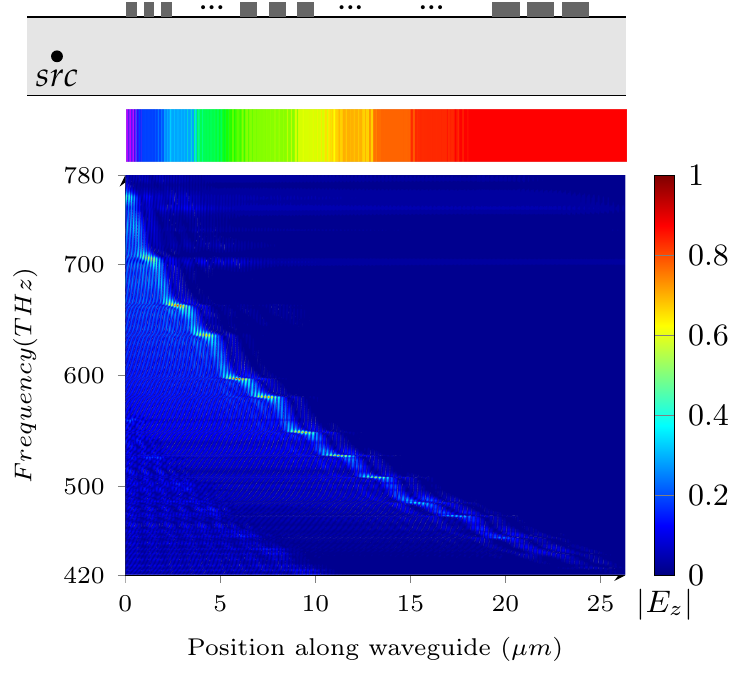}
\label{fig:DiscSpec}
}
\end{minipage}
\hfill
\begin{minipage}[c]{0.49\textwidth}
\subfloat[]{ %
\includegraphics[width=\textwidth]{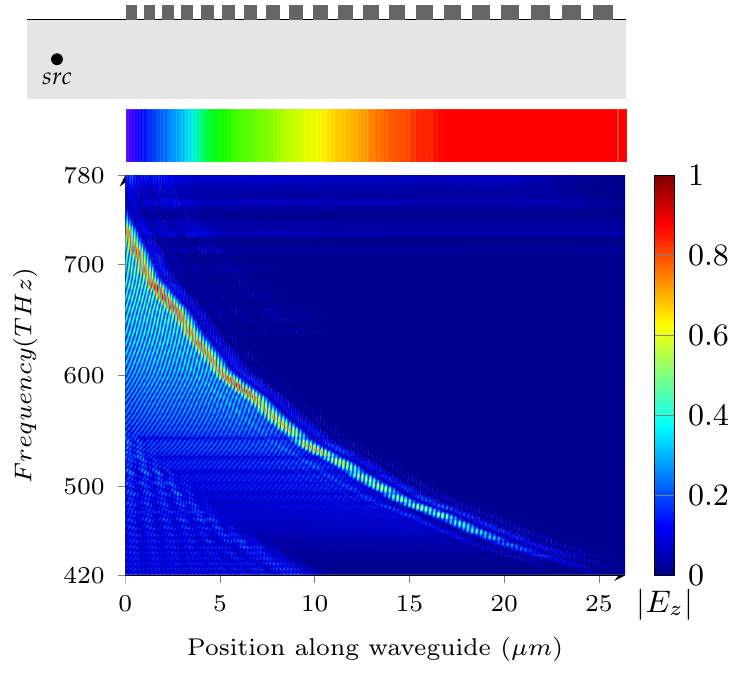}
\label{fig:ContSpec}
}
\end{minipage}
\caption{\protect\subref{fig:DiscSpec}: 
Top: The schematic of the rainbow-trapping device composed of multiple waveguide segments.
Bottom: 
Full wave analysis of the trapped rainbow obtain by FDTD simulation and Fast Fourier Transform. 
For the waveguide, $t=0.2 \mu m$. For the grating, $\Gamma=0.6$, $t=0.04 \mu m$, period range 
is designed from $0.13\to 0.26 \mu m$.
The Gaussian 
pulse enters from the left of the structure in the slab waveguide(not shown in the plot). 
Both figures demonstrate the field distribution in the frequency domain(vertical 
axis) and the spatial position(horizontal axis). 
\protect\subref{fig:ContSpec}: Rainbow trapping by waveguide of continuously varying parameters.  
The period varies from $0.130\mu m$ to $0.267\mu m$ gradually.  The
other design parameters and the excitation is the same as \protect\subref{fig:DiscSpec}.}
\end{figure}

To construct a trapped rainbow requires building a waveguide on which
the frozen mode is of different frequency at different position along the
device.  In our design, this is realized by placing multiple segments 
of the waveguide of different grating period one after another.  
A schematic is shown in the top of Fig.~\ref{fig:DiscSpec}.  
As a demonstration, our first device is composed of $14$ waveguide segments,
and the grating of each segment consists of 10 identical periods.  These 
gratings have the same thickness of $t = 40\text{nm}$ and duty cycle
$\Gamma = 0.6$, but the period varies from $130\text{nm}$ to $260\text{nm}$.  
The free-space wavelength of the frozen mode that would be supported by 
waveguides of these different designs range from
$395\text{nm}$ to $704\text{nm}$.  
The device is fed from the left by a slab waveguide made of the same material
and of the same thickness. A broad band pulse is launched in the feeding waveguide, and incident
to the trapping device from the left. ” 
We arrange the waveguide segments
so that the frozen mode frequency decreases from
left to the right, with the segments of higher frequency sitting 
at the upper stream of the optical power flow.  This is because, according to
Fig.~\ref{fig:disp} and Fig.~\ref{fig:disp_cmpr}, each segment actually supports
two modes of zero group velocity, one at the top of the lower band while 
the other at the bottom of the top band.  In our design, we use the lower band
of every segment, and the arrangement described above promises that the bottom of the 
upper band of each segment falls inside the bandgap of its neighboring upper stream segment,
thus would not be excited.  The structure is again simulated in MEEP.
In the simulation, we record the field at different 
positions along the center of the waveguide after the transient field 
fades out.  A Fourier transform then reveals the spectrum at each
position. The observed spectrum intensity at different positions along
the whole device is shown in the bottom of Fig.~\ref{fig:DiscSpec}. 
Here the horizontal axis is the lateral position along the device 
with the origin at the beginning of the first waveguide, while the
vertical axis is the signal frequency. The color shows the spectrum
intensity.  In the simulation, a pulse signal with approximately 
flat spectrum in the band of interests is used, so that no frequency
component has an advantage in the power intensity.
As we walk from left to the right along the device,
we can indeed observe 14 discrete steps in the spectrum 
at positions corresponding to the 14 waveguide segments
(the last one is less obvious due to the reflection by the segments ahead of it), 
going from $760\text{THz}$ (violet color) to $400\text{THz}$ (red color).
To give an intuitive understanding to the result, 
we show the color that would be observed at different positions along the
device rendered from the spectrum measured at the very position.  
The algorithm to render a color from a distribution of spectrum intensity is 
discussed in Ref.\cite{Spectra_Rendering}, and the result is shown in the 
middle of Fig.~\ref{fig:DiscSpec}.  The result clearly gives  a ``rainbow'' 
trapped along the device. The color shows the spectrum
intensity.  In the simulation, a pulse signal with approximately 
flat spectrum in the band of interests is used, so that no frequency
component has an advantage in the power intensity.
As we walk from left to the right along the device,
we can indeed observe 14 discrete steps in the spectrum 
at positions corresponding to the 14 waveguide segments
(the last one is less obvious due to the reflection by the segments ahead of it), 
going from $760\text{THz}$ (violet color) to $400\text{THz}$ (red color).
To give an intuitive understanding to the result, 
we show the color that would be observed at different positions along the
device rendered from the spectrum measured at the very position.  
The algorithm to render a color from a distribution of spectrum intensity is 
discussed in Ref.\cite{Spectra_Rendering}, and the result is shown in the 
middle of Fig.~\ref{fig:DiscSpec}.  The result clearly gives  a ``rainbow'' 
trapped along the device.
The former demonstration uses a piecewise continuous design.
To have a rainbow with adiabatic color change, we 
turn to a device with tapered design.  Rather than physically
tapering the thickness of the slab waveguide, we use 
continuously changed grating period along the whole device in the same range
as the former example, as we see in the
schematic shown on the top of Fig.~\ref{fig:ContSpec}.
A similar idea was used to make flat focusing lens in one of the authors' former works\cite{Fattal_Flat_2010a}. 
We expect the result to be a smooth-out version of the trapped rainbow
shown in Fig.~\ref{fig:DiscSpec}.  The result indeed proves our expectation
(refer to the middle of Fig.~\ref{fig:ContSpec}).  
As we see in the bottom of Fig.~\ref{fig:ContSpec},
the peak frequency of the spectrum changes as the observation position
changes, and a rainbow of continuously varying color can be observed.

\section{Conclusion}
In this paper we make use of the negative Goos-H\"anchen shift on the surface of
a dielectric grating to realize a frozen mode, i.e. a guided mode on a waveguide 
with no net power propagation.  Further, 
by tuning the design parameters of the grating on a waveguide, we can achieve frozen
modes with different frequencies sitting at different positions along the waveguide, 
so that a broad band pulse covering the whole visible spectrum 
can be caught by the waveguide, with different frequency components stored at different
positions.  The current design is, to the best of our knowledge, the first demonstration
of the ``trapped rainbow'' proposed in \cite{tsakmakidis_trapped_2007} that
make explicit use of the negative Goos-H\"anchen shift, the mechanism originally 
proposed in that paper. 
At the same time, the use of only dielectric materials promises a much lower ohmic 
loss.  The negative Goos-H\"anchen shift is realized on the surface of a grating, which
has a geometry much easier to be fabricated compared to the usually three dimensional
structure of a metamaterial.  Tuning the lateral design parameters rather than the
thickness further reduces the fabrication difficulty.  All these features make
the device suitable for practical use in areas such as slow light.

We should point out that the ``trapped rainbow'' serves as a manifesto of the 
capability of the grating in controlling the dispersion property of a slab 
waveguide.  According to \eqref{eqn:zgv_GHS}, the Goos-H\"anchen shift, or 
more generally, the reflection phase $\phi$ on the surface of the grating,
directly determines the behavior of the group velocity.  Since the reflection phase
is controlled by the design parameters of the grating, \eqref{eqn:zgv_GHS} 
gives us a straightforward method to synthesize the dispersion property of 
the waveguide as needed. 
We believe this dispersion engineering approach can have promising applications 
in optical networks. 
%
%
%



%


\end{document}